\documentclass{iau}

\usepackage{amsmath}
\usepackage{graphicx}
\usepackage{multirow}

\newcommand{\kms}{km~s$^{-1}$}

\newcommand{\doce}{\mbox{$^{12}$CO}}
\newcommand{\trece}{\mbox{$^{13}$CO}}
\newcommand{\diecisiete}{\mbox{C$^{17}$O}}
\newcommand{\dieciocho}{\mbox{C$^{18}$O}}
\newcommand{\hcop}{\mbox{HCO$^+$}}
\newcommand{\ndoshp}{\mbox{N$_2$H$^+$}}
\newcommand{\jno}{\mbox{$J$=9$-$8}}
\newcommand{\jss}{\mbox{$J$=7$-$6}}
\newcommand{\jsc}{\mbox{$J$=6$-$5}}
\newcommand{\jcc}{\mbox{$J$=5$-$4}}
\newcommand{\jct}{\mbox{$J$=4$-$3}}
\newcommand{\jtd}{\mbox{$J$=3$-$2}}
\newcommand{\jdu}{\mbox{$J$=2$-$1}}
\newcommand{\juc}{\mbox{$J$=1$-$0}}

\begin{document}

\lefttitle{M. Santander-Garc\'ia, E. Masa, J. Alcolea \& V. Bujarrabal}
\righttitle{Morpho-kinematical modelling in the molecular zoo beyond CO: the case of M~1-92}

\journaltitle{Planetary Nebulae: a Universal Toolbox in the Era of Precision Astrophysics}
\jnlDoiYr{2023}
\doival{10.1017/xxxxx}
\volno{384}

\aopheadtitle{Proceedings IAU Symposium}
\editors{O. De Marco, A. Zijlstra, R. Szczerba, eds.}
 
\title{Morpho-kinematical modelling in the molecular zoo beyond CO: the case of M~1-92
}

\author{Miguel Santander-Garc\'ia, Elisa Masa, Javier Alcolea, Valent\'in Bujarrabal}
\affiliation{Observatorio Astron\'omico Nacional (OAN-IGN), Spain}

\begin{abstract}
Ongoing improvements of sub-mm- and mm-range interferometers and single-dish radiotelescopes are progressively allowing the detailed study of planetary nebulae (PNe) in molecular species other than \doce\ and \trece. We are implementing a new set of tables for extending the capabilities of the morpho-kinematical modelling tool {\tt SHAPE+shapemol}, so radiative transfer in molecular species beyond \doce\ and \trece, namely \diecisiete, \dieciocho, HCN, HNC, CS, SiO, \hcop, and \ndoshp, are enabled under the Large Velocity Gradient approximation with the ease of use of {\tt SHAPE}. We present preliminary results on the simultaneous analysis of a plethora of IRAM-30m and HERSCHEL/HIFI spectra, and NOEMA maps of different species in the pre-PN nebula M~1-92, which show interesting features such as a previously undetected pair of polar, turbulent, high-temperature blobs, or a $^{17}$O/$^{18}$O isotopic ratio of 1.7, which indicates the AGB should have turned C-rich, as opposed to the apparent nature of its O-rich nebula.
\end{abstract}


\maketitle

\section{Introduction}

Over the last few decades, morpho-kinematical modelling of planetary nebulae (PNe) has provided a wealth of information useful for gaining insight into PN formation (e.g., \citealp{solf85,santander04, akashi13,akashi18}). Parameters derived from these models include the current 3D geometry and 3D velocity field of the ejecta, as well as its spatial orientation and kinematical (distance-dependent) age. These allow for instance to establish the evolutionary sequence of formation in the case of a ejecta composed of multiple structures, or investigating whether the equator of the PNe is aligned with the orbital plane of a close-binary central star (as models predict, see e.g. \citealp{jones20}), provided the latter is known.

Despite its proved usefulness, this kind of modelling has its limitations: it only provides a snapshot of the current state of the nebula, and it lacks information on other physical parameters necessary for hydrodynamic or magnetohydrodynamic modelling, such densities, temperatures and pressures at each location throughout the PN.

This shortcoming can be overcome, in the case of the molecular regime, by the inclusion of radiative transfer calculations of molecular species in these models. In this regard, the inclusion of {\tt shapemol} (\citealp{santander15}) in the user-friendly, successful morphokinematical modelling code {\tt SHAPE} (\citealp{steffen11}) enabled accurate non-LTE calculations of excitation and radiative transfer in \doce\ and \trece\ rotational lines under the Large Velocity Gradient (LVG) approximation (e.g. \citealp{castor70}), producing both synthetic single-dish profiles and spectral maps to match observations. Simultaneous fitting of lines of low- and high-excitation in this manner allows to derive the excitation conditions of the gas, namely its density and temperature, as well as the abundance of each species relative to that of molecular hydrogen. These in turn lead to estimates of the mass of the molecule-rich region of the PNe, provides some insight about the conditions for molecular chemistry, and further informs models of formation of PNe. 

We introduce the addition of several new molecular species in {\tt shapemol}, as well as a code to better allow for direct matching of interferometric observations. We use these tools to show preliminary results on the simultaneous analysis of a plethora of IRAM-30m and HERSCHEL/HIFI spectra, and NOEMA interferometric maps of different species in the pre-PN M~1-92.

The work presented here will be further discussed in much deeper detail by Masa et al. (in preparation).

\section{Expanding {\tt shapemol}}

The emission $j_\nu$ and absorption $k_\nu$ coefficients calculated by {\tt shapemol} for each molecular species and individual transition rely on pre-generated tables according to the density $n$ of each model grid cell, its temperature $T$, and the product $X\times\frac{r}{V}$, where $X$ is the abundance of the species with respect to H$_2$, and $r$ and $V$ are the distance to the central star and the expansion velocity, respectively. Tables provided for the previous version of {\tt shapemol} (\citealp{santander15}) spanned a range of physical parameters  expected to be found in the circumstellar environment of evolved low- and intermediate-mass stars, allowing to model ejecta with $n$ in the range 10$^2$ -- 10$^7$ cm$^{-3}$, $T$ from 10 to 1000 K, and $X\times\frac{r}{V}$ in the range $2.5\times10^{10}$ -- $1.3\times10^{13}$ cm~km$^{-1}$~s, and $8.3\times10^{8}$ -- $4.2\times10^{11}$ cm~km$^{-1}$~s for \doce\ and \trece\ respectively. These coverage was more focused on modelling of PNe, and were therefore insufficient for some projects. Modelling Asymptotic Giant Branch (AGB) stars, for instance, represented quite a struggle. Consequently, the new version of {\tt shapemol} considerably expands the table coverage for \doce\ and \trece\ by several orders of magnitude in $n$, and a broader range in $X\times\frac{r}{V}$. 

\begin{figure}
\begin{center}
 \includegraphics[width=10cm]{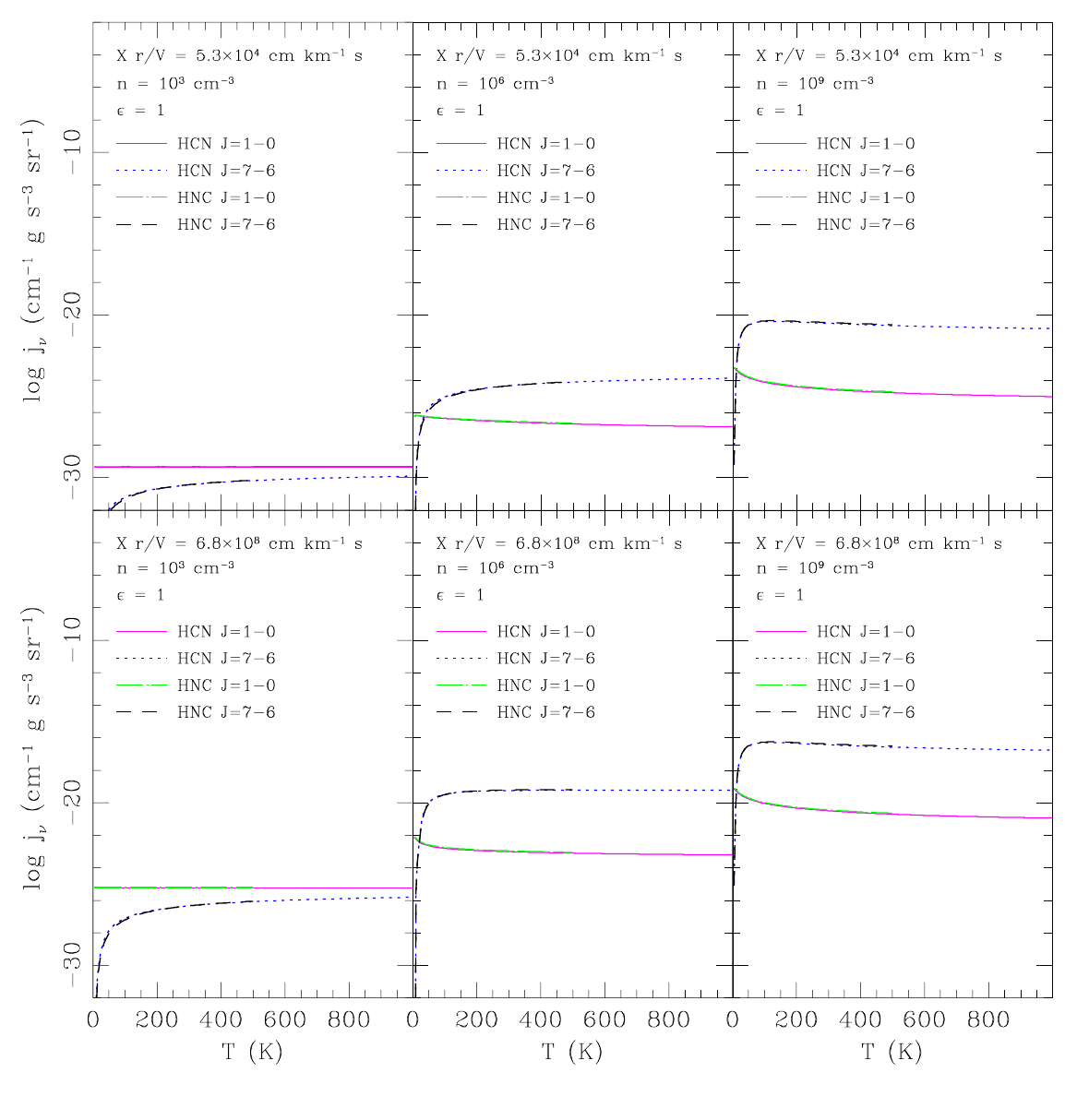}
\end{center}
\caption{Example of the emission coefficient $j_\nu$ of HCN and HNC under a range of conditions of abundance, density and temperature (and a unique value of the logarithmic velocity gradient, $\epsilon = dV/dr~r/V$, corresponding to ballistic expansion.}
\label{F1} 
\end{figure}

In addition, we have expanded the capabilities of {\tt shapemol} by including a whole new set of molecular species, namely \diecisiete, \dieciocho, HCN, HNC, CS, SiO, \hcop, and \ndoshp. The preliminary parameter coverage in these molecules, as well as the new coverage for \doce\ and \trece, can be found in Table~\ref{T1}. Note that the parameter coverage is susceptible to expand in the final version presented in Masa et al. (in preparation). An illustrative example of the emission coefficient $j_\nu$ of HCN and HNC for a broad range of physical conditions is shown in Figure~\ref{F1}.

The computed $j_\nu$ and $k_\nu$ were tested in two different manners. We first compared a simple, spherical model computed with {\tt SHAPE+shapemol} with theoretical calculations of the optical depth $\tau$ under LVG conditions, finding typical errors of a few \textperthousand\ in the optically thin case, increasing to a few \% as the nebula becomes optically thick. We also performed a battery of emission and absorption coefficients checks against the results provided by {\tt RADEX} (\citealp{vandertak07}). Although RADEX is not a LVG code, the small offsets we found (with typical values of $\sim$5\%) ensure our calculations are accurate enough.

{\fontsize{7}{9}\selectfont
\begin{table}[h!]
 \centering
 \caption{Molecules included and parameter coverage of the version of {\tt shapemol} presented in this work.}\label{T1}
 {\tablefont\begin{tabular}{@{\extracolsep{\fill}}lccc}
    \midrule
    Species & $n$ (cm$^{-3}$) & $T$ (K) & $X\times\frac{r}{V}$ (cm~km$^{-1}$~s) \\
    \midrule
    \doce   & 10$^2$ -- 10$^{12}$  & 5 -- 1000  &  $1.2\times10^{9}$ -- $6.1\times10^{13}$ \\
    \trece  & 10$^2$ -- 10$^{12}$  & 5 -- 1000  &  $3.8\times10^{7}$ -- $2.0\times10^{12}$ \\
    \diecisiete & 10$^2$ -- 10$^{12}$  & 5 -- 1000  &  $2.3\times10^{7}$ -- $1.2\times10^{12}$ \\
    \dieciocho &  10$^2$ -- 10$^{12}$  & 5 -- 1000  &  $1.2\times10^{7}$ -- $6.1\times10^{11}$ \\
    HCN     & 10$^2$ -- 10$^{12}$  & 5 -- 1000  &  $5.0\times10^{3}$ -- $5.0\times10^{10}$ \\
    HNC     & 10$^2$ -- 10$^{12}$  & 5 -- 1000  &  $1.7\times10^{3}$ -- $1.7\times10^{10}$ \\
    CS      & 10$^2$ -- 10$^{12}$  & 5 -- 1000  &  $2.0\times10^{4}$ -- $1.1\times10^{9}$ \\ 
    SiO     & 10$^2$ -- 10$^{12}$  & 5 -- 1000  &  $5.0\times10^{3}$ -- $2.7\times10^{8}$ \\
    \hcop   & 2.7$\times$10$^2$ -- 10$^{12}$$^\star$  & 5 -- 1000  &  $5.0\times10^{5}$ -- $2.7\times10^{10}$ \\
    \ndoshp & 2.7$\times$10$^2$ -- 10$^{12}$$^\star$  & 5 -- 1000  &  $5.0\times10^{2}$ -- $5.0\times10^{9}$ \\
    \midrule
    \end{tabular}}
\tabnote{\textit{Note}: The lower limit of the coverage marked by $\star$ is planned to expand in the final version provided in Masa et al. (in preparation).}  
\end{table}}

\section{Proper comparison of synthetic models and interferometric observations}

Interferometry involves emission filtering depending on the baselines and UV coverage. As a result, interferometric maps are not akin to Integral Field Spectroscopic data in the sense that they should not be compared with models in a direct fashion. Otherwise, flux lost in the observations (usually from extended, smooth structures) prevents a fair reconstruction by morpho-kinematical modelling.

Models simple enough (e.g. a flaring disk observed through an optical interferometer) can circumvent this limitation by direct comparison of model and observations in the UV plane. However, complex models with many parameters are a different story. In practice, comparisons are only possible in the image plane. Therefore, the model needs to undergo the same filtering and cleaning process suffered by the actual observations. This requires having accurate information of the interferometer configuration, baselines, etc., as well as of the source position in the sky at the moment(s) of observation. Essentially, we need the UV tables of the observation in order to apply the same filtering and cleaning process to the morpho-kinematical model. A proper comparison would also include the application of random noise similar to that of the observations to ensure an even better direct comparison.

We have devised a code for the radioastronomy {\tt GILDAS} suite\footnote{\url{https://www.iram.fr/IRAMFR/GILDAS}} that allows to quickly apply this process to models exported from {\tt SHAPE}, provided the UV table of the observation is available.

\section{Preliminary model of M~1-92}

In order to illustrate the new capabilities of {\tt SHAPE+shapemol} as well as the aforementioned {\tt GILDAS} code, we present a preliminary morpho-kinematical model of the pre-PN M~1-92. We use it to simultaneously fit NOEMA interferometric maps of the \jdu\ transition of \trece, \diecisiete, \dieciocho, HCN, and \hcop, as well as a large number of single-dish observations from different species made with the IRAM~30 radiotelescope and HERSCHEL/HIFI, namely \doce\ (\juc, \jdu, \jcc, \jss, \jno); \trece, \diecisiete\ and \dieciocho\ (\juc, \jdu); HNC and \ndoshp\ (\juc, \jtd); HCN and \hcop\ (\juc, \jdu, \jtd); CS (\jdu, \jtd); and SiO (\jdu, \jtd, \jct, \jcc, \jsc). We have kept the number of parameters down to a reasonable value in order to provide a comprehensive description of the geometry, motion, and physical conditions of the real ejecta.

\begin{figure}
\begin{center}
 \includegraphics[width=6.7cm]{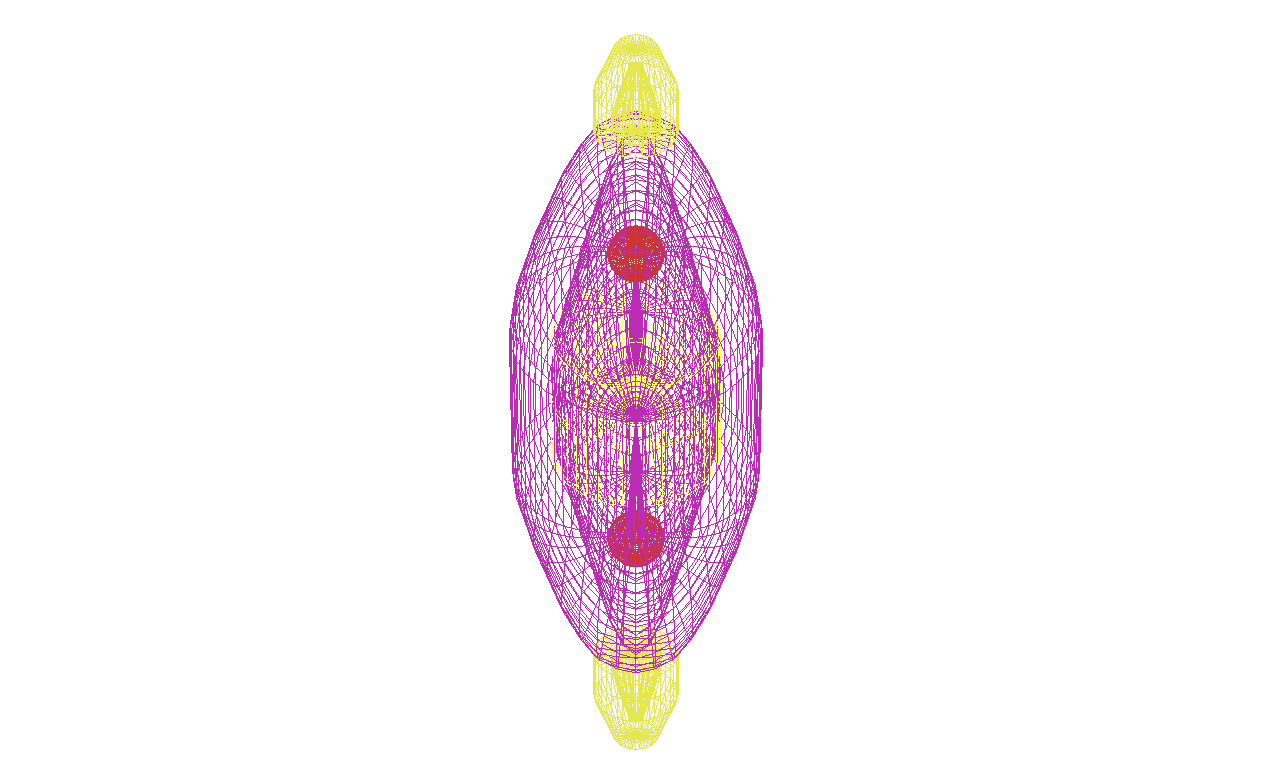}
 \includegraphics[width=6.7cm]{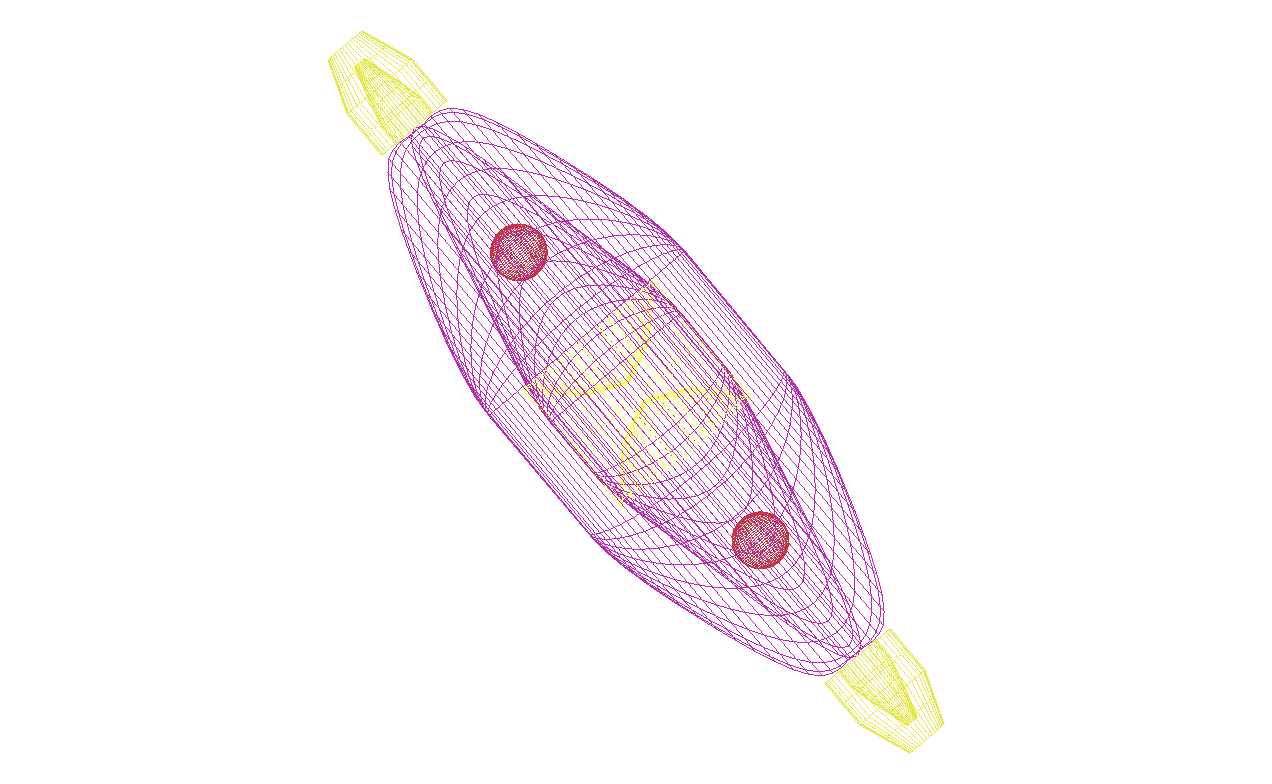}
\end{center}
\caption{Sketch of the preliminary morpho-kinematical model of M~1-92 presented in this work.}
\label{F2} 
\end{figure}

\begin{figure}
\begin{center}
 \includegraphics[width=13.5cm]{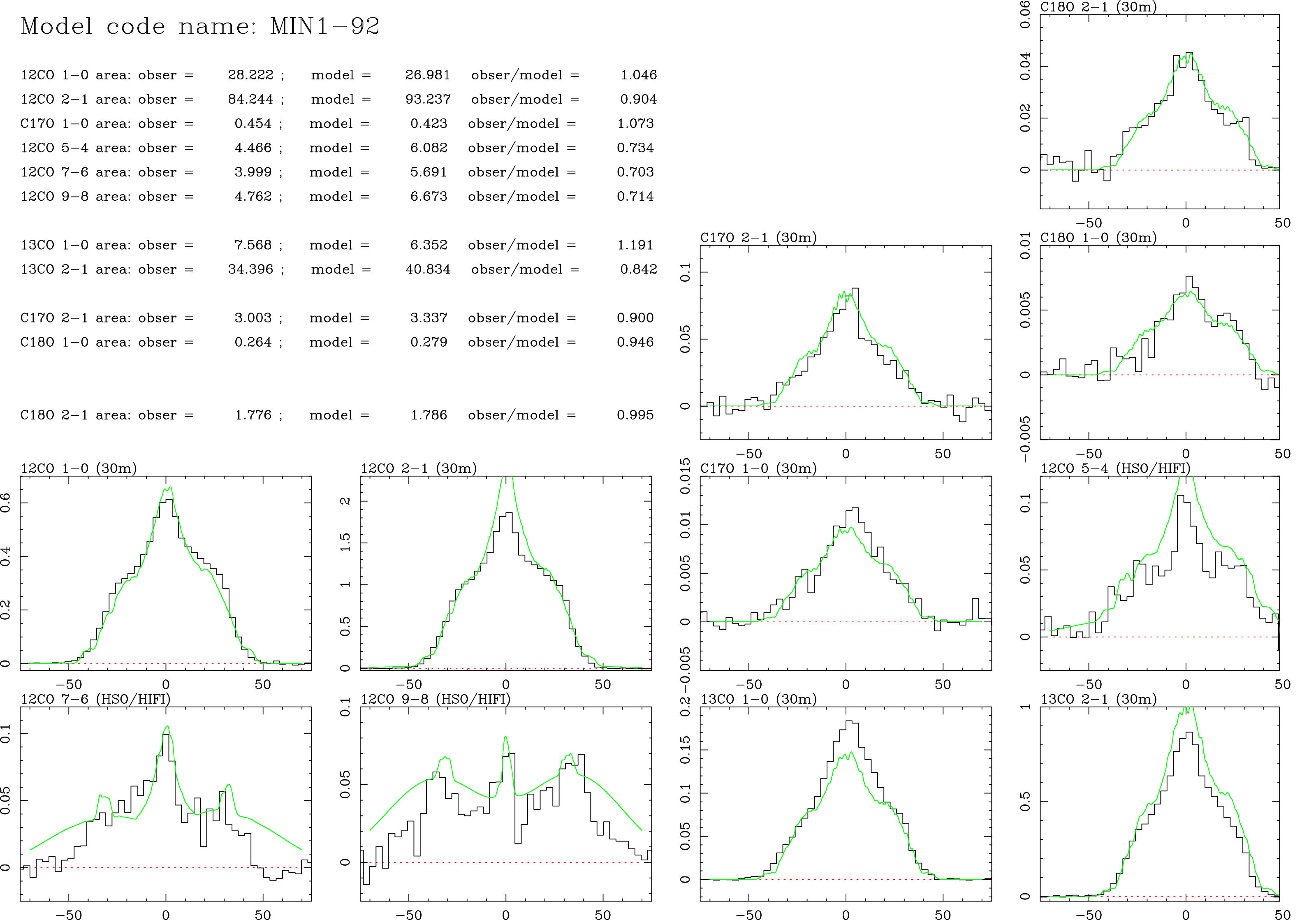}
\end{center}
\caption{Preliminary model fit to different single-dish spectral profiles from \doce, \trece, \diecisiete, and \dieciocho\ in M~1-92. Numbers in the top left section refer to integrated intensities of both the observation and model in each line, as well as their ratio in order to better assess the fairness of the fit.}
\label{F3} 
\end{figure}

\begin{figure}
\begin{center}
 \includegraphics[width=6.7cm]{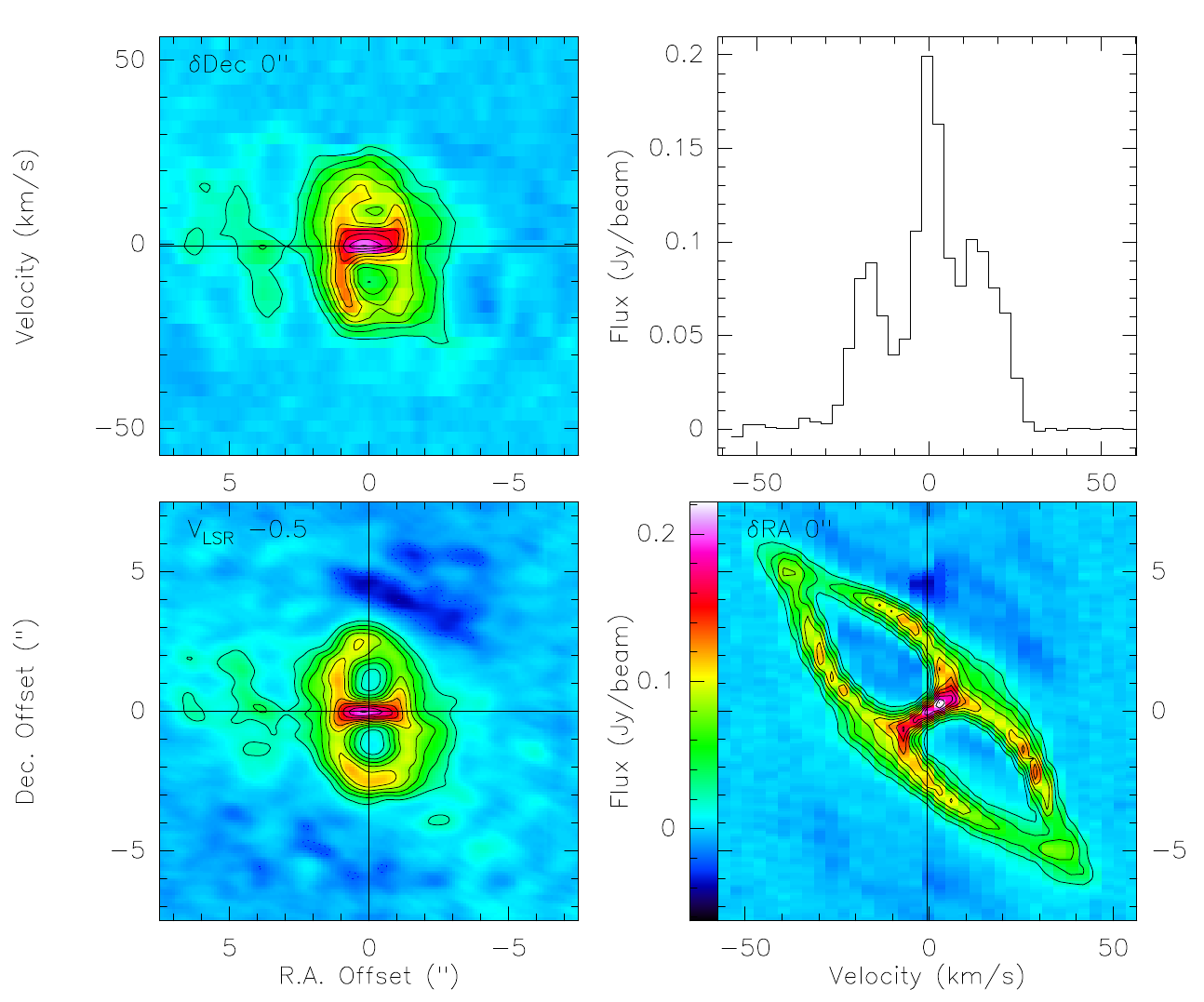}
 \includegraphics[width=6.7cm]{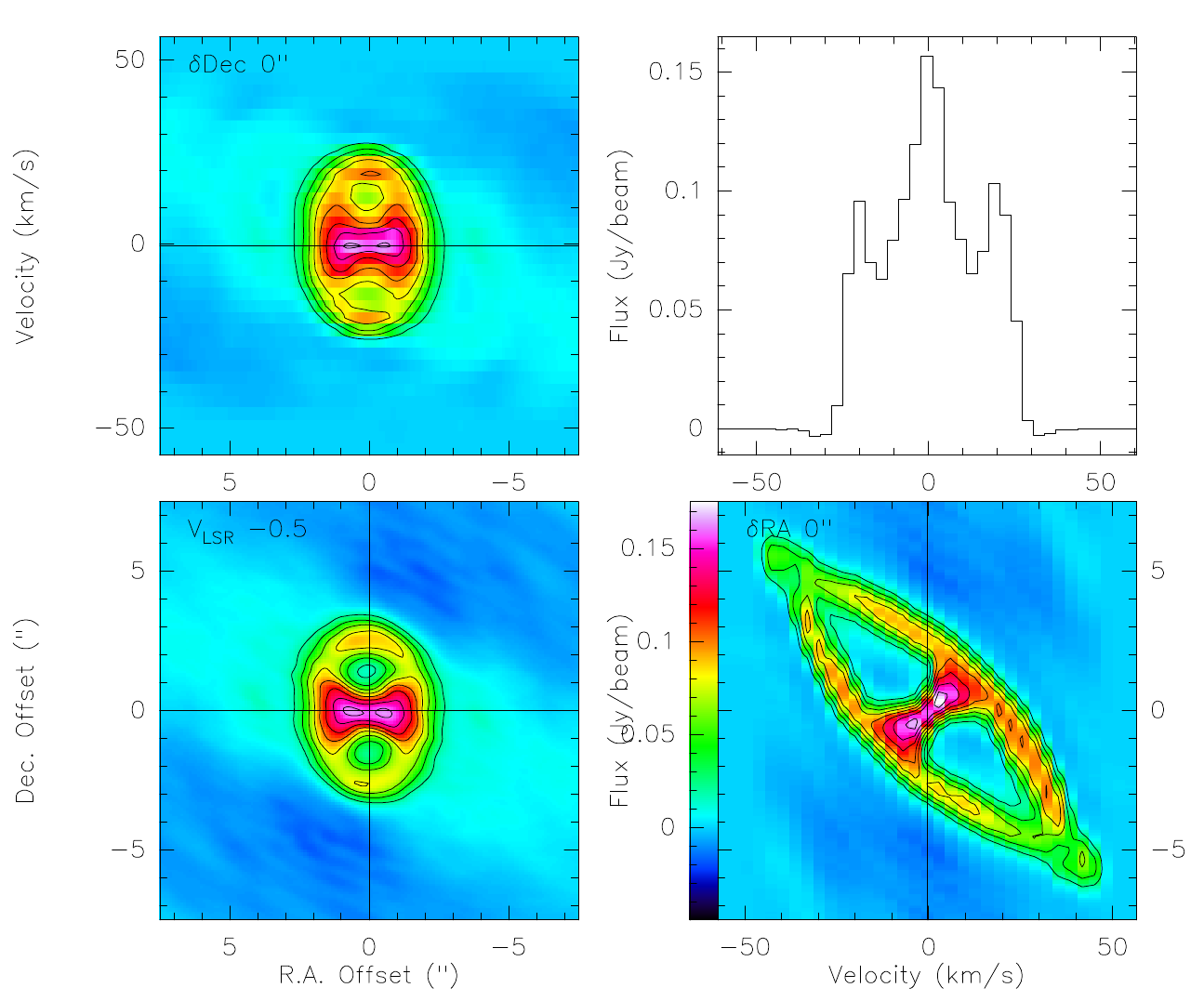}
\end{center}
\caption{\textit{Left:} A four-panel sample of the \trece\ \jdu\ NOEMA interferometric observations of the pre-PN M~1-92. The central (systemic velocity) channel of the spectral map is shown in the bottom left panel. The top left and bottom right panels show position velocity maps along the horizontal and vertical directions, respectively. The top right panel shows the spectral profile at the central position. \textit{Right:} Same for the preliminary model of M~1-92, after application of the GILDAS code to allow direct comparison with the observations.}
\label{F4} 
\end{figure}

\begin{figure}
\begin{center}
 \includegraphics[width=6.7cm]{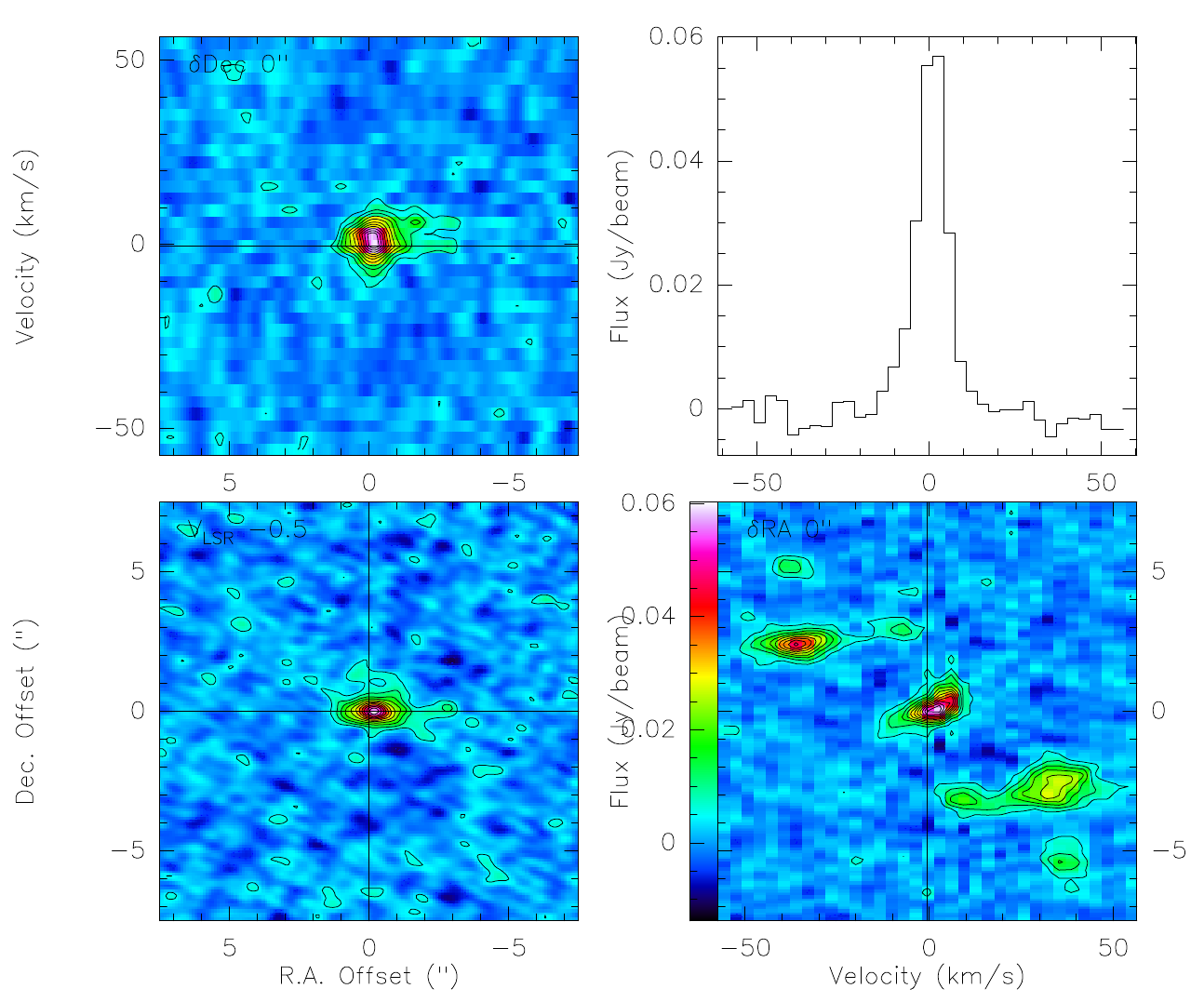}
 \includegraphics[width=6.7cm]{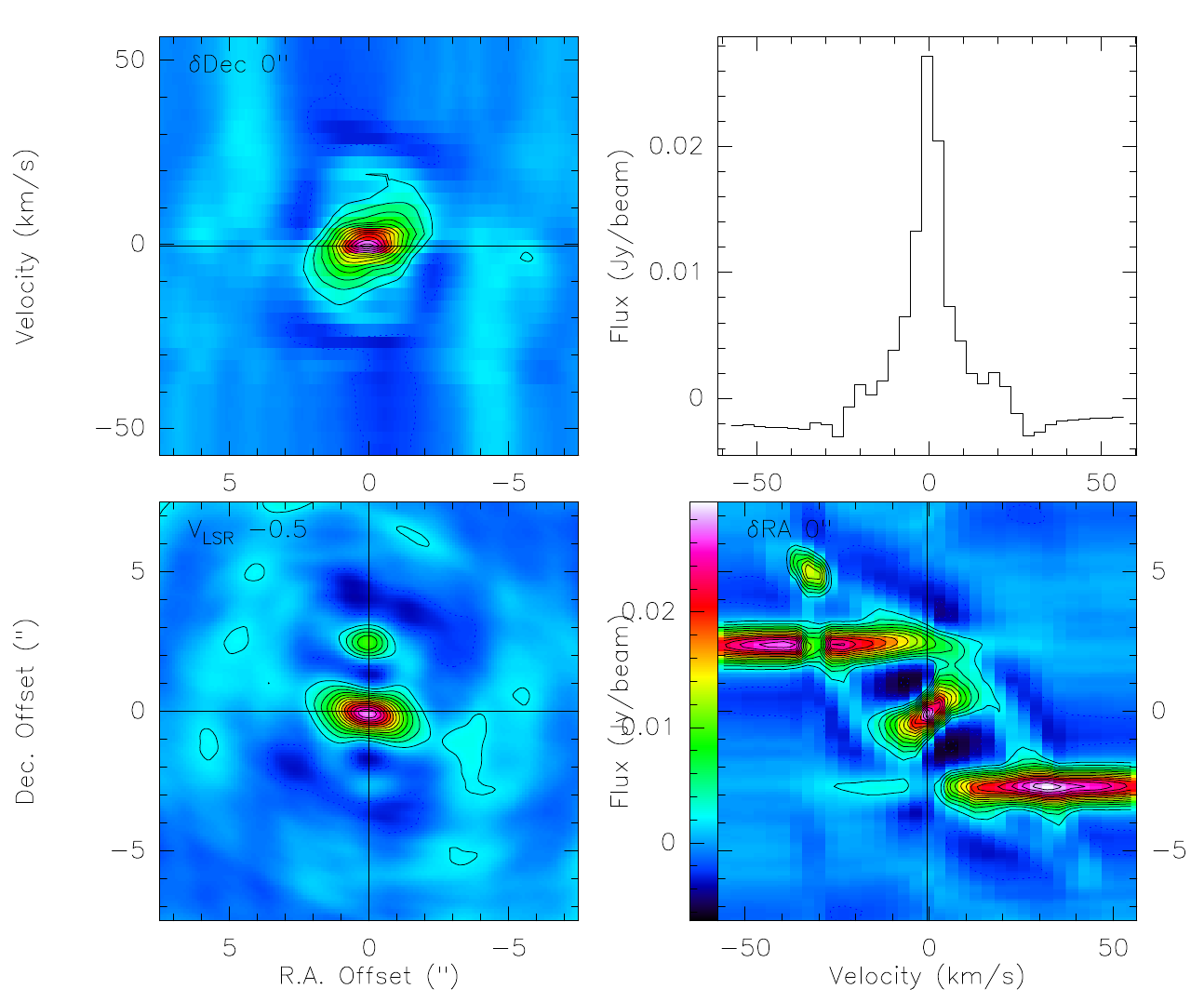}
\end{center}
\caption{\textit{Left:} A four-panel sample of the HCN \jdu\ NOEMA interferometric observations of the pre-PN M~1-92. The central (systemic velocity) channel of the spectral map is shown in the bottom left panel. The top left and bottom right panels show position velocity maps along the horizontal and vertical directions, respectively. The top right panel shows the spectral profile at the central position. \textit{Right:} Same for the preliminary model of M~1-92, after application of the GILDAS code to allow direct comparison with the observations.}
\label{F5} 
\end{figure}

A model sketch of this bipolar pre-PN is shown in Figure~\ref{F2}. A preliminary fit to different single-dish spectral profiles and interferometric maps can be seen in Figures~\ref{F3}, \ref{F4}, and \ref{F5}.

The modelling so far suggests that the tapered shape of the nebula is topped by two polar tips carved out by a hot polar outflow. The inner and outer regions of these tips have very different physical conditions. The density of the model ranges from 1.8$\times$10$^5$~cm$^{-3}$ in the equatorial inner region to a mere 10$^4$~cm$^{-3}$ in the inner region of the polar tips, which also are the hotter ones of the model, at $\sim$450~K, as opposed to the much colder walls of the nebula at 17~K. The nebula expands homologously, except for a couple of small, dense and hot blobs half way along the main axis which dash through the nebula at a much higher velocity (55~\kms) than the surrounding regions. These blobs, as well as the polar tips, are significantly bright in \hcop and HCN, but difficult to tell apart in low-$J$ \doce\ and \trece\ transitions.

In summary, the data suggests an active chemistry is taking place in shocked, hot regions within M~1-92. Additionally, an $^{17}$O/$^{18}$O isotopic ratio of 1.7 indicates the progenitor AGB should have turned carbon-rich during its evolution towards the tip of the AGB, whereas the resulting nebula is oxygen-rich. This discrepancy can be removed if the evolution of the AGB was suddenly interrupted by the event that gave rise to the current nebula in a sudden ejection, either by a close binary in common envelope or some other similar phenomenon.

A full description of the final version of this model will be given in Masa et al. (in preparation).

\bigskip

We acknowledge funding from the Spanish Ministry of Science and Innovation (MICINN) through project EVENTs/Nebulae-Web (grant PID2019-105203GB-C21).

\vfill\eject

\end{document}